\documentclass{article}

\usepackage[english]{babel}

\usepackage[a4paper,top=2cm,bottom=2cm,left=3cm,right=3cm,marginparwidth=1.75cm]{geometry}

\usepackage{amssymb}
\usepackage{siunitx}
\PassOptionsToPackage{hyphens}{url}\usepackage{hyperref}
\usepackage{cleveref}
\usepackage[utf8]{inputenc}
\usepackage{csquotes}
\usepackage{booktabs}
\usepackage{longtable}
\usepackage{adjustbox}
\usepackage{array}
\usepackage{url}
\usepackage{titlesec}
\usepackage{authblk}
\usepackage{xcolor} 

\newcommand{\bn}{\begin{equation}}
\newcommand{\en}{\end{equation}}
\newcommand{\ba}{\begin{array}{ll}}
\newcommand{\ea}{\end{array}}

\titleformat{\subsection}
  {\mdseries\itshape\large} 
  {\thesubsection}{1em}{} 

\usepackage[english]{babel}
\usepackage{subcaption}

\crefformat{figure}{#2Figure~#1#3}
\Crefformat{figure}{#2Figure~#1#3}
\crefformat{table}{#2Table~#1#3}
\Crefformat{table}{#2Table~#1#3}
\crefformat{section}{#2Section~#1#3}
\Crefformat{section}{#2Section~#1#3}

\author{
    \large YiCheng Dai\footnote{\href{mailto:YichengDai@mail.ecust.edu.cn}{YichengDai@mail.ecust.edu.cn}},  ~
    \large Wei Liao\footnote{\href{mailto:liaow@mail.ecust.edu.cn}{liaow@ecust.edu.cn}} \\
    \vspace{3mm}
    \large School of Physics, East China University of Science and Technology, \\
    130 Meilong Road, Shanghai 200237, P. R. China
}

\title{Reionization History and Neutrino Mass}

\begin{document}
\date{}
\maketitle

\begin{abstract} 
Recent  baryon acoustic oscillation (BAO) distance measurements, when combined with Cosmic Microwave Background (CMB) observations in the $\Lambda$CDM framework, lead to a preference for negative neutrino masses.
We investigate whether this neutrino mass anomaly can be alleviated by a class of astrophysically motivated reionization histories.
Using a frequentist analysis, we find that some reionization histories can move the best-fit value of $\sum m_\nu$ to a positive value and bring $\sum m_\nu\simeq0.06~{\rm eV}$ into the 95\% confidence interval.
To separate the effect of the total optical depth from that of the details of the reionization history, 
we compare a high-$\tau$ history with a two-step tanh-like reionization history of the same $\tau$.
The resulting $\Delta\chi^2(\sum m_\nu)$ profiles are nearly identical.
This indicates that the effect is mainly driven by the total optical depth, while the details of the reionization history play only a minor role.
\end{abstract}

\section{Introduction}
Neutrinos are the lightest known massive fermions and play an important role in both particle physics and cosmology.
However, the absolute values of their masses  remain unknown.
Solar neutrino oscillation experiments can only determine $\Delta m_{21}^2 \equiv m_2^2-m_1^2$,
which is the mass squared difference between the mass eigenstates $\nu_2$ and $\nu_1$.
Atmospheric neutrino oscillation experiments can only determine the absolute of the other splitting, 
$|\Delta m_{32}^2|$, which involves the third mass eigenstate $\nu_3$.
These measurements imply a lower bound on the total neutrino mass \cite{Esteban_2017, Gonzalez_Garcia_2021},
\[
\sum m_\nu \gtrsim 0.06~\mathrm{eV}
\]
for normal ordering $\Delta m_{32}^2>0$, and
\[
\sum m_\nu \gtrsim 0.10~\mathrm{eV}
\]
for inverted ordering $\Delta m_{32}^2<0$.

Since oscillation experiments do not determine the absolute mass scale, independent probes are needed.
Cosmological observations provide one of the most sensitive probes of the absolute scale of the neutrino mass.
In particular, the combination of CMB observations and BAO distance measurements provides a strong constraint on $\sum m_\nu$ through both the background evolution and the growth of structure.

However, a recent Bayesian analysis combining BAO measurements from Dark Energy Spectroscopic Instrument (DESI) Data Release 1 with CMB data from Planck and the Data Release 6 of Atacama Cosmology Telescope (ACT) has obtained a very stringent upper limit on the neutrino mass, $\sum m_\nu < 0.072~\mathrm{eV}$ at 95\% confidence level, within the $\Lambda$CDM model and assuming three degenerate neutrino masses \cite{Adame_2025}.
Since this limit is very close to the lower bound implied by neutrino oscillation experiments for normal ordering, it raises a possible tension between cosmology and oscillation data \cite{elbers2025constraintsneutrinophysicsdesi, PhysRevD.111.063534, Craig:2024tky}.
Beyond Bayesian analyses, this preference for low neutrino masses is also found in a frequentist analysis using DESI data\cite{chebat2025cosmologicalneutrinomassfrequentist}.
The likelihood as a function of the sum of neutrino masses $\Delta\chi^2(\sum m_\nu)$ is obtained by minimizing the $\chi^2$ function for a fixed value of $\sum m_\nu$.
The likelihood profile is then fitted with a quadratic function near its minimum and extrapolated into the unphysical region $\sum m_\nu<0$.
The fitted minima lie in this negative mass region and indicate an anomalous preference for an unphysical value of $\sum m_\nu$ within the $\Lambda$CDM framework.

Several possibilities have been studied to understand or alleviate this tension.
One direction is to focus on different data combinations, and to test how the preference depends on the BAO distance measurements or on the treatment of the CMB likelihoods \cite{chebat2025cosmologicalneutrinomassfrequentist,DESI:2024mwx, descollaboration2026darkenergysurveyyear, Calabrese_2025, PhysRevD.110.123537}.
Another direction is to extend the cosmological model by introducing dynamical dark energy or new physics in the neutrino or the dark sector \cite{Ahlen_2025}.

A more conservative way is to revisit the role of $\tau$, the optical depth of reionization.
Previous studies have shown that a larger reionization optical depth, or a relaxation of the low-$\ell$ EE constraint on $\tau$, can alleviate the CMB+BAO tension with $\sum m_\nu\simeq0.06~\mathrm{eV}$ \cite{6vd2-rbfn}.
However, previous studies mostly treated $\tau$ as a free parameter or imposed an external prior on it.
It remains unclear whether astrophysically allowed reionization histories can yield an optical depth large enough to alleviate the negative neutrino mass preference in CMB+BAO fits.
It is also important to determine whether this effect is controlled by the total optical depth or by the details of the reionization history.

In this work, we explore this possibility by using several astrophysically allowed ionization histories, $x_e(z)$, as inputs to the theoretical calculations and the subsequent CMB+BAO fits.
We then compute the frequentist profile likelihood as a function of  $\sum m_\nu$ for each history.
We further compare different histories at fixed optical depth to separate the effect of $\tau$ from that of the detailed reionization shape.

The rest of this paper is organized as follows.
Sec.~\ref{Sec.2} discusses the role of the reionization optical depth and introduces the reionization history models used in this work.
The data set and the frequentist profile likelihood method are described in Sec.~\ref{Sec.3}.
Sec.~\ref{Sec.4} presents our frequentist results and discusses the effects of the reionization history and the optical depth on the neutrino mass preference.
Finally, Sec.~\ref{Sec.5} summarizes the main results of this work.

\section{Optical Depth, Reionization History and Neutrino Mass}
\label{Sec.2}
In this section, we first show how a larger reionization optical depth $\tau$ can alleviate the neutrino mass tension.
We then discuss the role of reionization history in cosmological fits.

Recent analyses of CMB and BAO data have shown that the preference for negative neutrino mass is highly sensitive to the reionization optical depth \cite{6vd2-rbfn}.
High-$\ell$ CMB temperature and polarization spectra mainly constrain the combination $A_s e^{-2\tau}$, where $A_s$ is the amplitude of the primordial scalar curvature perturbations.
Therefore, a larger $\tau$ needs a larger value of $A_s$ to keep the CMB amplitude fixed.
A larger $A_s$  can enhance the structure growth and CMB lensing which, on the other hand, can be suppressed by a larger neutrino mass.
These two effects can compensate each other and can lead to a positive correlation of larger $\tau$ and larger neutrino mass in cosmological fits.
As a result, a higher $\tau$ is possible to relax the preference for a negative neutrino mass to a positive value.

Although high-$\ell$ CMB data cannot break the degeneracy between $\tau$ and $A_s$,
introducing low-$\ell$ EE power spectrum data together with  the high-$\ell$ CMB data can remove the degeneracy and yield a constraint on $\tau$.
However, given that the relevant polarization signal is more than two orders of magnitude weaker than the temperature anisotropies, the resulting constraint on $\tau$ is sensitive to foreground modeling and instrumental systematics.
Current analyses find that $\tau\simeq0.09$ is in tension with the Planck low-$\ell$ EE measurement at about the $2.7\sigma$ level ~\cite{6vd2-rbfn}.
Although such a high optical depth is not preferred by the present low-$\ell$ EE data, but the level of disagreement is still moderate.
On the other hand,  $\tau$ can also be determined by the reionization history.
This motivates us to examine whether astrophysically allowed reionization histories can yield a sufficiently large optical depth and how their predicted low-$\ell$ EE spectra affect the neutrino mass preference.

\begin{figure}[htbp]
\centering
\includegraphics[width=1\textwidth]{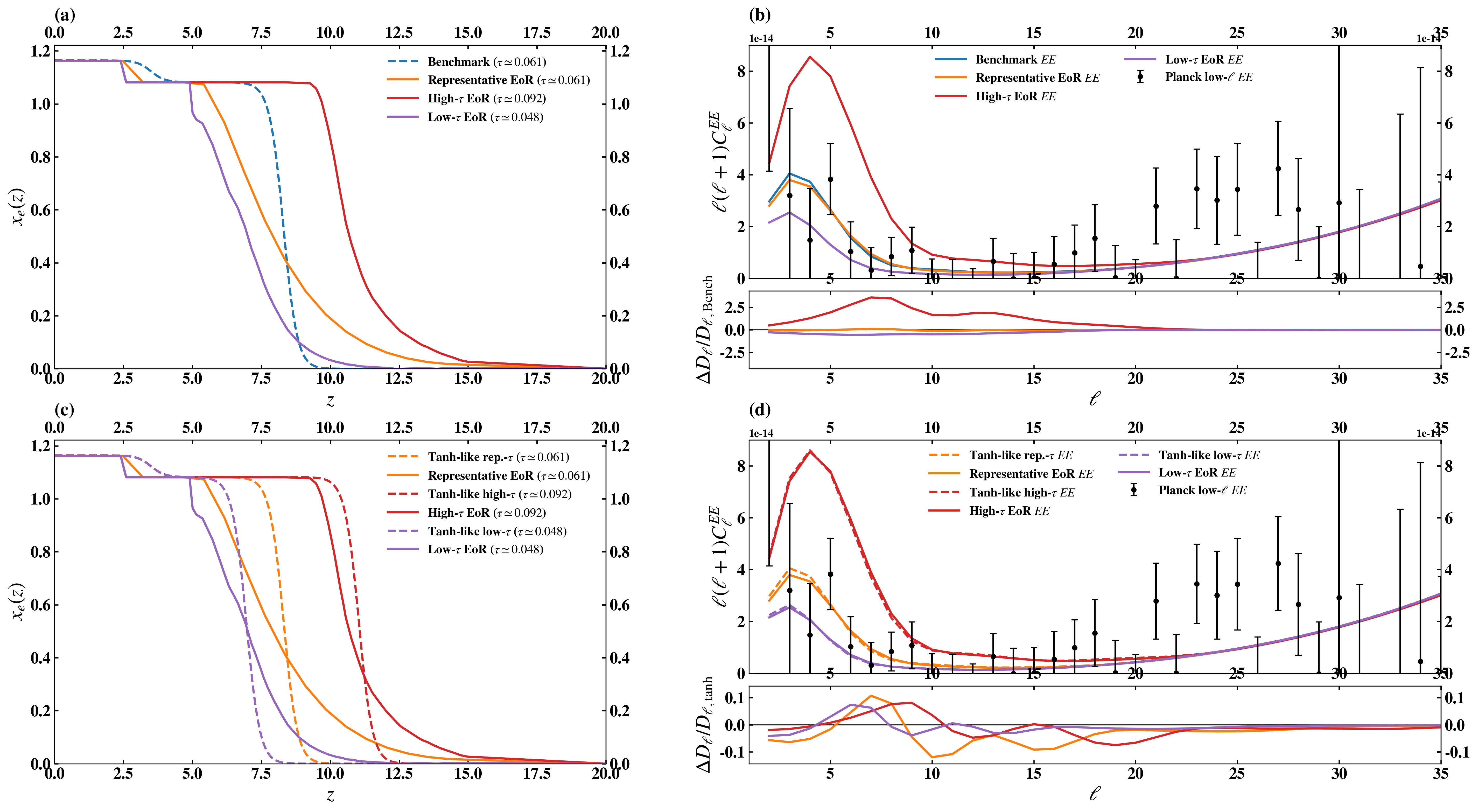}
    \caption{
Reionization histories and the corresponding low-$\ell$ EE power spectra.
In panel (a), the solid lines show the three astrophysically allowed EoR histories: Low-$\tau$ EoR, Representative EoR, and High-$\tau$ EoR, 
while the dashed blue line is the benchmark model evaluated at $\tau\simeq0.061$. 
(b) The colored lines show the low-$\ell$ EE power spectra corresponding to the reionization histories in panel (a).
The black points and error bars are the low-$\ell$ EE data from Planck 2018 \cite{Planck2018EEdata}.
The lower sub-panel shows the relative difference with respect to the benchmark history, defined as
$\Delta D_\ell^{EE}/D_{\ell,\rm Bench}^{EE} \equiv
(D_{\ell}^{EE}-D_{\ell,\rm Bench}^{EE})/D_{\ell,\rm Bench}^{EE}$, where $D_\ell^{EE}\equiv \ell(\ell+1)C_\ell^{EE}$.
(c) The solid lines show the same EoR histories as in panel (a).
The dashed lines show the tanh-like history profiles with the same optical depths corresponding to the three astrophysically allowed EoR histories
shown in panel (a).
The same color denotes the same optical depth.
This comparison is used to isolate the effect of the detailed reionization history.
(d) shows the low-$\ell$ EE spectra corresponding to the reionization histories in panel (c), using the same line styles and colors.
The solid and dashed curves with the same color nearly overlap.
The lower sub-panel shows the relative difference at fixed optical depth.  
The cosmological parameters used to generate the curves in this figure are listed in Appendix~\ref{app:fig1_params}.
}
    \label{fig: Reionization_histories_and_lowlEE}
\end{figure}

We introduce one benchmark reionization history and three astrophysically allowed histories, as shown in the upper left panel of Fig.~\ref{fig: Reionization_histories_and_lowlEE}.
The benchmark case is the fiducial reionization model implemented in CLASS \cite{DiegoBlas2011}, with $\tau$ left as a free parameter.
For illustration, the blue dashed curve in the upper left panel of Fig.~\ref{fig: Reionization_histories_and_lowlEE} shows this model for the particular choice $\tau\simeq0.061$.
It uses the default tanh-like reionization parametrization in CLASS \cite{lesgourgues2011cosmiclinearanisotropysolving}, which contains one tanh transition for hydrogen reionization and another tanh transition for the second ionization of helium at low redshift.
The other three history profiles are taken from Ref.~\cite{Qin_Mesinger_2025} and are denoted as Low-$\tau$ EoR, Representative EoR, and High-$\tau$ EoR, according to their resulting optical depths.
They are shown by the purple, red, and orange solid lines in the upper left panel of Fig.~\ref{fig: Reionization_histories_and_lowlEE}, respectively.
These histories describe different redshift evolutions of the free-electron fraction $x_e(z)$ during the epoch of reionization.
Their main difference is the amount and redshift distribution of free electrons along the line of sight.
A history with more free electrons during reionization gives a larger optical depth.
Therefore, for these three cases, $\tau$ is not an independent parameter, but is derived from the specified $x_e(z)$, with only a mild dependence on the background cosmological parameters.
The Representative EoR history corresponds to the red curve in Fig.~4 in Ref.~\cite{Qin_Mesinger_2025}.
The Low-$\tau$ and High-$\tau$ EoR histories are constructed from the upper and lower edges of the light-blue 95\% confidence region in the same figure, respectively.
We note that Fig.~4 of Ref.~\cite{Qin_Mesinger_2025} shows the neutral fraction $\bar{x}_{\rm HI}$ of  the intergalactic medium (IGM), while our calculation uses the normalized free-electron fraction $x_e(z)=n_e/n_{\rm H}$.
In converting $\bar{x}_{\rm HI}$ to $x_e(z)$, we include the contribution from helium ionization.
We also include the second ionization of helium at low redshift, so that 
$\lim_{z\rightarrow 0}x_e(z)\simeq 1.16$.
Therefore, $x_e(z)$ is not simply equal to $1-\bar{x}_{\rm HI}$.
Since $\bar{x}_{\rm HI}$ and $x_e(z)$ are inversely related, the upper edge of the allowed region in $\bar{x}_{\rm HI}$ corresponds to the Low-$\tau$ EoR history, while the lower edge corresponds to the High-$\tau$ EoR history.
This light-blue 95\% confidence region is obtained using the UV luminosity functions and the CMB optical depth constraint, but does not include the Lyman-$\alpha$ forest effective optical depth distributions \cite{Qin_Mesinger_2025}.
Thus, the phrase ''astrophysically allowed'' here refers only to the constraints from UV luminosity functions and the CMB optical depth.

The upper right panel of Fig.~\ref{fig: Reionization_histories_and_lowlEE} shows  the low-$\ell$ EE power spectra derived for
these four reionization history profiles.
The low-$\ell$ EE power spectrum from Planck 2018 is also plotted for comparison.
Different reionization histories lead to visible differences in the low-$\ell$ EE power spectrum, especially around the reionization bump at $\ell\lesssim10$.
In particular, the High-$\tau$ EoR history produces the largest bump amplitude, while the Low-$\tau$ EoR history produces the smallest one, consistent with their different optical depths.

The four history profiles differ not only in their total optical depths, but also in the redshift dependence of the ionization fraction $x_e(z)$.
To separate these two effects, we construct a varied history profile for each astrophysical EoR history by using the benchmark tanh-like profile and 
by fixing $\tau$ to the same value. 
These varied history profiles are shown in the lower left panel of Fig.~\ref{fig: Reionization_histories_and_lowlEE}.
As shown in the lower right panel of Fig.~\ref{fig: Reionization_histories_and_lowlEE}, although the detailed reionization histories are different, the resulting low-$\ell$ EE power spectra become very similar once the optical depth is fixed.
The maximum relative difference among these three pairs is below $10\%$.
Moreover, as noted in Ref.~\cite{Planck:2018legacy}, the large scale CMB polarization signal is much weaker than the temperature anisotropy signal on the same angular scales, and, as a result, the current low-$\ell$ EE measurement is sensitive to systematics and foregrounds.
Therefore, the reionization bump is not expected to provide a strong discrimination among different detailed reionization histories in cosmological fits, especially when the optical depth is fixed.
This suggests that, for the history profiles considered here, the low-$\ell$ EE spectra are controlled mainly by the total optical depth.
We therefore expect that the alleviation of the negative neutrino mass tension, if present, is driven mainly by the value of $\tau$, rather than by the details of the reionization history itself.
We will explore this point in detail in the cosmological fit in the following section of the present article.

\section{Methodology and Data}
\label{Sec.3}
In this section, we briefly describe the cosmological model, the dataset, and the frequentist analysis method used in this study.

We work in a flat $\Lambda$CDM+$\sum m_\nu$ framework.
In our study, the varied cosmological parameters include the physical baryon density $\omega_b\equiv\Omega_bh^2$, the physical cold dark matter density $\omega_{\rm cdm}\equiv\Omega_{\rm cdm}h^2$, the reduced Hubble parameter $h$, the scalar amplitude $A_s$ varied as $\ln(10^{10}A_s)$, and the scalar spectral index $n_s$. 
The neutrino sector is described by three degenerate massive neutrinos, whose total mass is denoted by $\sum m_\nu$.

The treatment of the optical depth $\tau$ is slightly different for different reionization histories.
For tabulated EoR histories, $\tau$ is not varied independently, but is determined by the input ionization fraction $x_e(z)$.
For the benchmark case, by contrast, we use the default tanh-like reionization history and $\tau$ is treated as a free parameter.
In the following analysis we consider five history profiles, i.e.,  the benchmark reionization model,  three astrophysically allowed EoR history profiles defined in Sec.~\ref{Sec.2}, and the tanh-like high-$\tau$ profile introduced in Sec.~\ref{Sec.2}.
The tanh-like high-$\tau$ profile has a two-step tanh-like reionization shape but with the optical depth fixed to $\tau=0.092$, the same value of
the high-$\tau$ EoR profile.
The tanh-like high-$\tau$ is used to separate the effect of the total optical depth from that of the detailed reionization shape.
For clarity, the reionization history profiles used in the following analysis are summarized in the first column of Table~\ref{tab:profile_summary}.

In all analyses of different reionization histories, we use the same CMB+BAO data combination.
The CMB data include the Planck 2018 PR3 \cite{Planck2018results} low-$\ell$ TT and EE likelihoods, the Planck 2018 plik high-$\ell$  TTTEEE likelihoods, and the CMB lensing likelihoods from Planck PR4 \cite{Carron_2022}, ACT DR6 \cite{Qu_2024, Qu_2026}, and SPT-3G MUSE \cite{Ge_2025}.
For BAO data, we use the DESI DR2 BAO distance measurements \cite{elbers2025constraintsneutrinophysicsdesi, Abdul_Karim_2025}.
For these likelihoods, we include the recommended nuisance parameters and use broad flat priors for the cosmological parameters.

The theoretical models are computed with the Boltzmann code CLASS.
The main frequentist analysis in this study is performed by Procoli \cite{Karwal:2024qpt}, a code for extracting the 1D profile likelihood of a parameter using the MontePython likelihood interface \cite{Brinckmann:2018cvx, Audren:2012wb}.
We label the reionization cases in Table~\ref{tab:profile_summary} by $i$.
For each case $i$ and each fixed value of $\sum m_\nu$, Procoli minimizes the total $\chi^2$ with respect to the remaining cosmological and nuisance parameters.
This gives the profiled $\chi^2$ curve
\[
\chi^2_{i,{\rm prof}}(\sum m_\nu).
\]
We then define
\[
\Delta\chi^2_i(\sum m_\nu)
=
\chi^2_{i,{\rm prof}}(\sum m_\nu)
-
\chi^2_{i,{\rm prof,min}},
\]
where $\chi^2_{i,{\rm prof,min}}$ is the minimum value among the sampled profile points for the same reionization case $i$.
The resulting $\Delta\chi^2_i(\sum m_\nu)$ profile is then fitted with a quadratic function near its minimum.
For cases where the profile extends beyond $\Delta\chi^2=4$, we perform two fits using points with $\Delta\chi^2<4$ and $\Delta\chi^2<9$, respectively.
The  the best-fit value  $\sum m_\nu^\mathrm{bf}$ is allowed to lie in the unphysical region, $\sum m_\nu<0$, after the quadratic extrapolation.
Since the best-fit value depends mildly on the chosen fitting range, we report both  the two best fits in Table~\ref{tab:profile_summary}.
For cases where all displayed points remain within $\Delta\chi^2<4$, a single quadratic fit is used.

\section{Results}
\label{Sec.4}
In this section, we present our fitting results and discuss the impact of the optical depth and the detailed reionization history on 
the likelihood profile as a function of  $\sum m_\nu$.

\begin{figure}[htbp]
\centering
\includegraphics[width=1\textwidth]{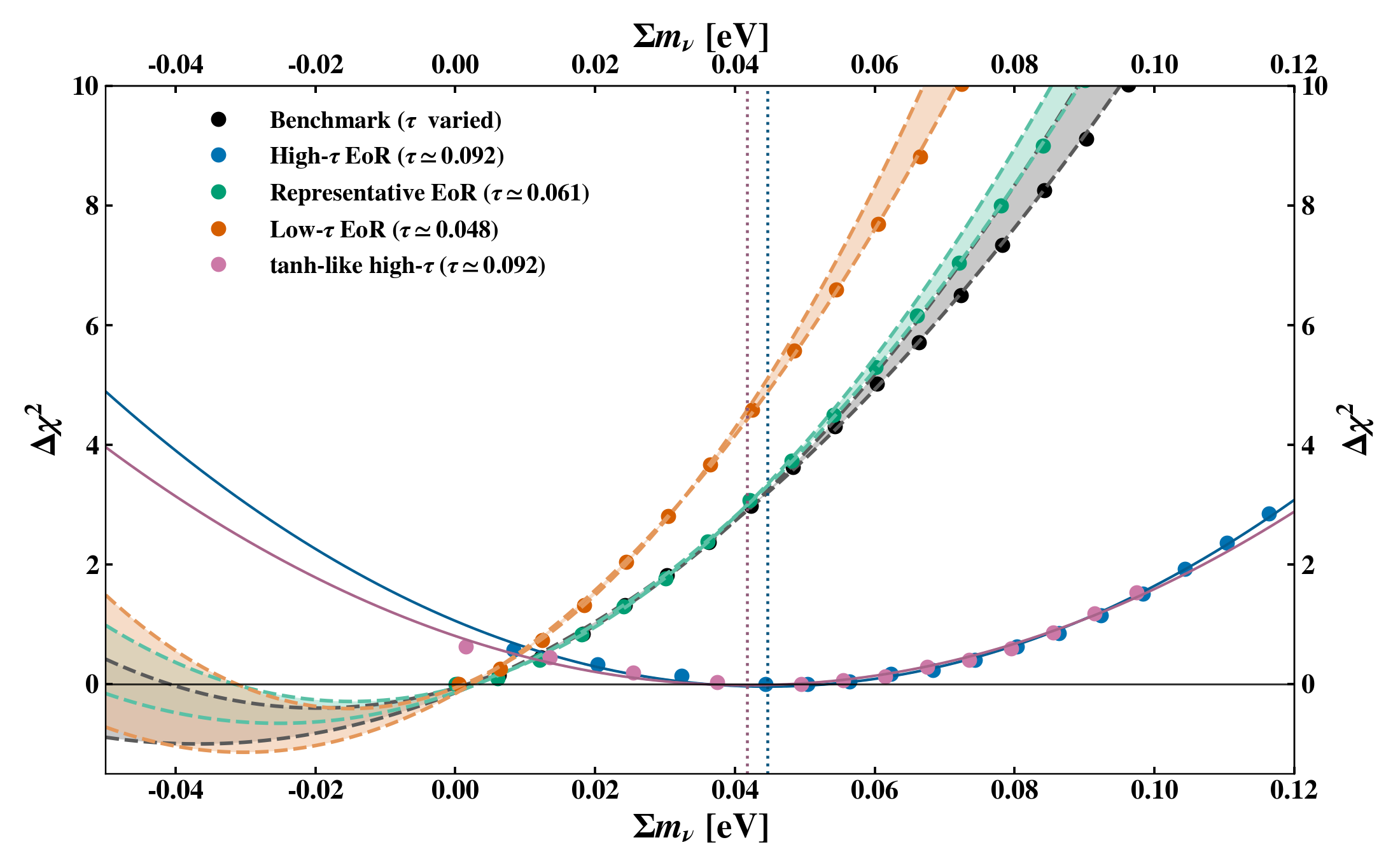}
    \caption{
Frequentist $\Delta\chi_i^2(\sum m_\nu)$ profiles of the total neutrino mass for different reionization histories.
The dots show the calculated profile points.
For the benchmark, Representative EoR and Low-$\tau$ EoR cases, the dashed lines are the quadratic fitting results obtained using the profile points with $\Delta\chi^2<4$ and $\Delta\chi^2<9$, and the shaded regions show the range between these two fits.
For the benchmark case, $\tau$ is varied along the profile.
Its best-fit values range from approximately $0.061$ to $0.064$ for the sampled points with $\Delta\chi^2<4$, and from $0.064$ to $0.068$ for the other points with $4<\Delta\chi^2<9$.
For the High-$\tau$ EoR and tanh-like high-$\tau$ cases, we perform only one quadratic fit because all displayed profile points lie within $\Delta\chi^2<4$.
The blue dotted vertical line marks  fitted minimum $\sum m_\nu^\mathrm{bf}$ of the quadratic fit for High-$\tau$ EoR profile.
The pink dotted vertical line marks  fitted minimum $\sum m_\nu^\mathrm{bf}$ of the quadratic fit for  tanh-like high-$\tau$ profile.
}
    \label{fig: Fit_profile}
\end{figure}

The main frequentist results are shown in Fig.~\ref{fig: Fit_profile}.
The black dots, gray curves, and gray band show the frequentist results for the benchmark case.
The dots represent the $\Delta\chi_i^2$ profile, while the two gray dashed curves show the corresponding quadratic fits obtained using the points with $\Delta\chi^2<4$ and $\Delta\chi^2<9$, respectively.
Both fits reach their minima at negative values of $\Sigma m_\nu$, namely $-0.0196~\mathrm{eV}$ and $-0.0368~\mathrm{eV}$.
In this benchmark case, $\tau$ is treated as a free parameter in the fitting.
Thus, along the $\Delta\chi^2(\Sigma m_\nu)$ profile, $\tau$ takes different best-fit values ranging from 0.061 to 0.068 approximately.
The green results correspond to the Representative EoR model, whose derived optical depth, $\tau\approx0.061$, is close to the value of $\tau$ at one 
profile point in the benchmark case. Its $\Delta\chi^2$ profile is very close to that of the benchmark case, and the two quadratic fits give minima at $-0.0148~\mathrm{eV}$ and $-0.0253~\mathrm{eV}$.
The orange results correspond to the Low-$\tau$ EoR model, whose optical depth is smaller, $\tau\approx0.048$.
Although the positions of the fitted minima, $-0.0150~\mathrm{eV}$ and $-0.0302~\mathrm{eV}$, are close to those of the benchmark and Representative EoR cases, the overall profile is visibly different.
Nevertheless, both of the positions of the fitted minima remain in the negative mass region, and show a preference for negative values of $\Sigma m_\nu$.

The situation changes once the High-$\tau$ EoR model, which corresponds to a much larger optical depth $\tau\approx0.092$, is taken into account.
The blue dots show the $\Delta \chi_{\mathrm{High-}\tau~\mathrm{EoR}}^2$ profile, while the blue curve shows the quadratic fit obtained using all the profile points.
Since there are no points with $\Delta \chi^2>4$, no additional fit or band is shown in this case.
The blue dotted vertical line marks the position of minimum of this fit.
In this case, the preference for negative neutrino mass is significantly alleviated, since the position of the fitted minimum shifts from the negative region to a positive value, $\Sigma m_\nu = 0.0447~\mathrm{eV}$.
This indicates that a sufficiently large optical depth can alleviate the preference for negative neutrino mass.
Although the fitted minimum remains slightly below the minimal physical value at $\Sigma m_\nu\simeq 0.06~\mathrm{eV}$, this physical value lies within the profile-extrapolated 68\% interval as shown in the High-$\tau$ EoR row of Table~\ref{tab:profile_summary}.
Therefore, the High-$\tau$ EoR case is compatible with $\Sigma m_\nu\simeq 0.06~\mathrm{eV}$.

To further separate the effect of the optical depth from that of the detailed reionization history, we also consider the tanh-like reionization model, with $\tau=0.092$ fixed.
The results are shown by the pink dots and curve.
As shown in Fig.~\ref{fig: Fit_profile}, the quadratic fitting minimum is at $\Sigma m_\nu=0.0417~\mathrm{eV}$, very close to the value obtained in the High-$\tau$ EoR case, $\Sigma m_\nu=0.0447~\mathrm{eV}$.
Moreover, the overall shape of the fitted curve is nearly identical.
This comparison suggests that, for the high-$\tau$ reionization histories considered here, changing the details of reionization history has only a minor impact on the $\Delta \chi^2$ profile for neutrino mass once the optical depth is fixed.
We note that this conclusion should remain valid for high-$\tau$ reionization histories similar to those considered here.
Histories with significantly different shapes  might lead to different results.

The  best fit $\sum m_\nu^\mathrm{bf}$ alone does not fully characterize the impact of reionization history on the neutrino mass constraint.
We also need to check whether the physical value implied by neutrino oscillation experiments, $\Sigma m_\nu\simeq 0.06~\mathrm{eV}$, lies below the profile-extrapolated  68\% upper bound or  95\% upper bound.
These upper bounds are obtained from the fitted profile by solving
\[
\Delta\chi^2(\Sigma m_\nu)-\Delta\chi^2(\Sigma m_\nu^{\rm bf})=1 ~\mathrm{or} ~3.84 .
\]

\begin{table}[htbp]
\centering
\renewcommand{\arraystretch}{1.22}
\caption{
Summary of the frequentist results for different reionization histories.
The columns $\Sigma m_{\nu,68}^{\rm upper}$ and $\Sigma m_{\nu,95}^{\rm upper}$ denote the upper bounds of the profile-extrapolated 68\% and 95\% intervals, obtained from the fitted neutrino-mass profile by solving
$\Delta\chi^2(\Sigma m_\nu)-\Delta\chi^2(\Sigma m_\nu^\mathrm{bf})=1$ and $3.84$, respectively, where $\Sigma m_\nu^\mathrm{bf}$ is the position of the minimum of the fitted quadratic curve.
For cases with $\Sigma m_\nu^\mathrm{bf}<0$, this minimum is an extrapolated value.
For the Benchmark, Representative EoR, and Low-$\tau$ EoR cases, the two rows correspond to quadratic fits using profile points with $\Delta\chi^2<4$ and $\Delta\chi^2<9$, respectively.
}
\label{tab:profile_summary}
\begin{tabular}{llcccc}
\toprule
Reionization case $i$
& Fit range
& $\tau$
& $\Sigma m_\nu^{\rm bf}$ [eV]
& $\Sigma m_{\nu,68}^{\rm upper}$ [eV]
& $\Sigma m_{\nu,95}^{\rm upper}$ [eV]
\\
\midrule

Benchmark
& $\Delta\chi^2<4$
& 0.061-0.064
& $-0.0196$
& $0.0141$
& 0.0464 \\

& $\Delta\chi^2<9$
& 0.061-0.068
& $-0.0368$
& $0.0030$
& 0.0411 \\

Low-$\tau$ EoR
& $\Delta\chi^2<4$
& 0.048
& $-0.0150$
& $0.0104$
&0.0348 \\

& $\Delta\chi^2<9$
& 0.048
& $-0.0302$
& $0.0003$
& 0.0296 \\

Representative EoR
& $\Delta\chi^2<4$
& $0.061$
& $-0.0148$
& $0.0164$
& 0.0463 \\

& $\Delta\chi^2<9$
& $0.061$
& $-0.0253$
& $0.0098$
& 0.0435 \\

High-$\tau$ EoR
& single fit
& 0.092
& $0.0447$
& $0.0873$
& 0.1282 \\

tanh-like high-$\tau$
& single fit
& 0.092
& $0.0417$
& $0.0877$
& 0.1318 \\

\bottomrule
\end{tabular}
\end{table}

For clarity, Table~\ref{tab:profile_summary} lists the fitted positions of minima and the corresponding profile-extrapolated 68\% and 95\% upper bounds for the different reionization cases.
The $\Sigma m_{\nu,68}^{\rm upper}$ column shows that $\Sigma m_\nu\simeq 0.06~\mathrm{eV}$ is not within the 95\% interval in the first three cases, which have relatively low optical depths.
On the other hand,   $\Sigma m_\nu\simeq 0.06~\mathrm{eV}$  is within the 68\% interval  in the last two high-$\tau$ cases.
Therefore, a high optical depth not only moves the best-fit value to a positive value, but also makes $\Sigma m_\nu\simeq 0.06~\mathrm{eV}$ compatible with the profile-extrapolated 68\% interval, and hence also the 95\% interval.
The similarity between the High-$\tau$ EoR and the tanh-like high-$\tau$ cases further supports that this effect is mainly driven by the optical depth rather than by the detailed reionization history.

\section{Conclusion and Discussions}
\label{Sec.5}
In this work, we studied whether astrophysically allowed reionization histories can alleviate the preference for negative neutrino mass in $\Lambda$CDM fits to CMB+BAO data.
We implemented several reionization histories and computed the frequentist profile likelihood of $\sum m_\nu$.

We find that the High-$\tau$ EoR history gives a positive best-fit value of $\sum m_\nu$, and brings $\sum m_\nu\simeq0.06~\mathrm{eV}$ into the profile-extrapolated 68\% confidence range.
In contrast, benchmark reionization, representative EoR and the low-$\tau$ EoR histories still prefer negative values of $\sum m_\nu$, and the normal ordering lower bound implied by neutrino oscillation experiments $\sum m_\nu\simeq0.06~\mathrm{eV}$ still lies above the corresponding profile-extrapolated 95\% upper bounds.
This study suggests that more precise measurements of the optical depth can be very helpful to clarify the neutrino mass tension and are well motivated.

We further compare the High-$\tau$ EoR history with the tanh-like high-$\tau$ case, which has the same optical depth but uses the benchmark two-step tanh-like reionization shape.
The two cases give nearly identical $\Delta\chi^2(\sum m_\nu)$ profiles.
This shows that, for the histories considered here, the change in the profile likelihood of $\sum m_\nu$ is mainly driven by the total optical depth $\tau$, rather than by the details of the reionization history.

We also note several limitations of this analysis.
First, the Low-$\tau$ EoR and High-$\tau$ EoR histories used in this work are constructed from the light blue allowed region in Fig.~4 of Ref.~\cite{Qin_Mesinger_2025}.
We choose this region because it allows a wider range of optical depths, including histories with relatively large $\tau$, and therefore makes the impact on the neutrino mass profile easier to identify.
However, this region is obtained using only the UV luminosity functions and the CMB optical-depth constraint, without including the Lyman-$\alpha$ forest effective optical depth distributions.
Therefore, the phrase ''astrophysically allowed'' in this work should be understood in this restricted sense, and the high-$\tau$ histories considered here should be only regarded as phenomenological test cases.

Moreover, the reionization histories considered here are relatively short, and do not include reionization histories with a high redshift tail.
Due to the computational cost, we have mainly performed the fixed-$\tau$ comparison at $\tau\simeq0.092$.
A broader scan over optical depths and reionization shapes is left for future work.

\bigskip
\section*{Note}
\label{sec:Note}

During the preparation of this work, we became aware of Ref.~\cite{6r54-8lv4}, which also emphasized the role of a larger reionization optical depth in alleviating the neutrino mass tension.
Our research is different from that in Ref.~\cite{6r54-8lv4} in two aspects.
Ref.~\cite{6r54-8lv4} mainly investigated the cosmological impact of increasing $\tau$, for example by replacing the low-$\ell$ polarization constraint with external priors on $\tau$.
In contrast, we focus on the effect of astrophysically motivated reionization history profiles and study whether optical depths compatible with
astrophysical observations can relax the preference of negative neutrino mass to positive values in cosmological fits.
We further compare effects of reionization history profiles of the same optical depth in order to separate the effects of the total $\tau$ from that of the detailed profile of $x_e(z)$.

\bigskip
\section*{Acknowledgements}
\label{sec:acknowledgements}
W. Liao is supported by National Natural Science Foundation of China under the grant No. 11875130.

\appendix

\section{Cosmological parameters used in Fig.~\ref{fig: Reionization_histories_and_lowlEE}}
\label{app:fig1_params}

Table~\ref{tab:fig1_cosmo_params} lists the cosmological parameters used to generate the reionization histories and the corresponding low-$\ell$ EE spectra shown in Fig.~\ref{fig: Reionization_histories_and_lowlEE}. 
Rather than choosing ad hoc parameter values for this illustrative figure, we use the parameters associated with the profile analysis.
For the benchmark model, the three tabulated EoR histories, and the tanh-like high-$\tau$ case, the cosmological parameters are taken from the sampled point where the corresponding profile $\Delta\chi^2(\sum m_\nu)$ in Fig.~\ref{fig: Fit_profile} reaches its minimum.
For the benchmark model, $\tau$ is varied in the profile analysis, and the benchmark curve in Fig.~\ref{fig: Reionization_histories_and_lowlEE} is evaluated at $\tau\simeq0.061$, the value corresponding to the sampled minimum point.
For the tanh-like high-$\tau$ case, $\tau$ is fixed to $0.092$.
For the three tabulated EoR histories, $\tau$ is computed from the specified reionization history and the background cosmological parameters.
For the tanh-like rep-$\tau$ and tanh-like low-$\tau$ histories, we use the same background cosmological parameters as the corresponding tabulated EoR case, but replace the reionization history by the tanh-like form with the same optical depth.

\begin{table}[htbp]
\centering
\renewcommand{\arraystretch}{1.15}
\caption{
Cosmological parameters used to generate the reionization histories and low-$\ell$ EE spectra shown in Fig.~\ref{fig: Reionization_histories_and_lowlEE}.
}
\label{tab:fig1_cosmo_params}
\resizebox{\textwidth}{!}{
\begin{tabular}{llcccccc}
\toprule
Case
& Reionization model
& $h$
& $\omega_b$
& $\omega_{\rm cdm}$
& $\ln(10^{10}A_s)$
& $n_s$
& $\sum m_\nu~[\mathrm{eV}]$
\\
\midrule

Benchmark
& tanh-like, $\tau\simeq0.061$
& 0.6858
& 0.02251
& 0.11845
& 3.0555
& 0.9705
& $2.76\times10^{-4}$
\\

Representative EoR
& tabulated $x_e(z)$
& 0.6858
& 0.02252
& 0.11850
& 3.0560
& 0.9706
& $8.26\times10^{-5}$
\\

High-$\tau$ EoR
& tabulated $x_e(z)$
& 0.6882
& 0.02259
& 0.11717
& 3.1093
& 0.9746
& $4.43\times10^{-2}$
\\

Low-$\tau$ EoR
& tabulated $x_e(z)$
& 0.6840
& 0.02248
& 0.11889
& 3.0330
& 0.9686
& $4.56\times10^{-4}$
\\

Tanh-like rep.-$\tau$
& tanh-like, $\tau=\tau_{\mathrm{Rep.}\text{-}\tau\,\mathrm{EoR}}$
& 0.6858
& 0.02252
& 0.11850
& 3.0560
& 0.9706
& $8.26\times10^{-5}$
\\

Tanh-like high-$\tau$
& tanh-like, $\tau=0.092$
& 0.6884
& 0.02262
& 0.11709
& 3.1088
& 0.9752
& $4.95\times10^{-2}$
\\

Tanh-like low-$\tau$
& tanh-like, $\tau=\tau_{\mathrm{Low}\text{-}\tau\,\mathrm{EoR}}$
& 0.6840
& 0.02248
& 0.11889
& 3.0330
& 0.9686
& $4.56\times10^{-4}$
\\

\bottomrule
\end{tabular}
}
\end{table}

\end{document}